\documentclass[epj]{webofcMy} % my (slightly tuned) version of the style-file 
\usepackage[utf8]{inputenc}
\usepackage[varg]{txfonts}    % Web of Conferences font
\usepackage{booktabs}
\usepackage{xcolor}
\usepackage{mathtools} % includes amsmath
\definecolor{darkred}{rgb}{0.4,0.0,0.0}
\definecolor{darkgreen}{rgb}{0.0,0.4,0.0}
\definecolor{darkblue}{rgb}{0.0,0.0,0.4}
\usepackage[bookmarks,linktocpage,colorlinks,
            linkcolor=darkred,
            urlcolor =darkblue,
            citecolor=darkgreen]{hyperref}
\usepackage{subfigure}
\usepackage[justification=justified,
            labelfont=bf,
            font=small,
            singlelinecheck=off]{caption} % enforce justification
\wocname{EPJ Web of Conferences}
\woctitle{Lattice2017}
\usepackage{macros}
\begin{document}
%%%%%%%%%%%%%%%%%%%%%%%%%%%%%%%%%%%%%%%%%%%%%%%%%%%%%%%%%%%%%%%%%%%%%%%%%%%%%
%
%%%%%%%%%%%%
% title page
%%%%%%%%%%%%
%
\selectlanguage{english}
%----------------------------------------------------------------------------
\hfill\parbox{14mm}{\texttt{%     % put preprint numbers here
MS-TP-17-17
}}
%----------------------------------------------------------------------------
\title{%
Non-perturbative determination of $\cv$, $\zv$ and $\zs/\zp$\\
in $\nf=3$ lattice QCD\thanks{%
Talk at the 35th International Symposium on Lattice Field Theory
(LATTICE 2017), 18-24 June 2017, Granada, Spain
}
}
%----------------------------------------------------------------------------
\author{%
\firstname{Jochen}
\lastname{Heitger}
\inst{1}\fnsep\thanks{Speaker, \email{heitger@uni-muenster.de}} \and
\firstname{Fabian} \lastname{Joswig}\inst{1} \and
\firstname{Anastassios} \lastname{Vladikas}\inst{2} \and
\firstname{Christian} \lastname{Wittemeier}\inst{1}
}
%----------------------------------------------------------------------------
\institute{%
Westf\"alische Wilhelms-Universit\"at M\"unster, 
Institut f\"ur Theoretische Physik,\\
Wilhelm-Klemm-Stra{\ss}e 9, 48149 M\"unster, Germany
\and
INFN, Sezione di Tor Vergata,
c/o Dipartimento di Fisica, Universit\`{a} di Roma Tor Vergata,\\
Via della Ricerca Scientifica 1, 00133 Rome, Italy
}
%----------------------------------------------------------------------------
\abstract{%
We report on non-perturbative computations of the improvement coefficient
$\cv$ and the renormalization factor $\zv$ of the vector current in
three-flavour $\Or(a)$ improved lattice QCD with Wilson quarks and 
tree-level Symanzik improved gauge action.
To reduce finite quark mass effects, our improvement and normalization 
conditions exploit massive chiral Ward identities formulated in the 
Schrödinger functional setup, which also allow deriving a new method to 
extract the ratio $\zs/\zp$ of scalar to pseudoscalar renormalization 
constants.
We present preliminary results of a numerical evaluation of $\zv$ and $\cv$
along a line of constant physics with gauge couplings corresponding to
lattice spacings of about $0.09\,\Fm$ and below, relevant for
phenomenological applications.
}
%----------------------------------------------------------------------------
\maketitle
%----------------------------------------------------------------------------
%
%%%%%%%%%%%%
% paper body
%%%%%%%%%%%%
%
%----------------------------------------------------------------------------
\section{Introduction}
\label{sec:intro}
%----------------------------------------------------------------------------
%
A popular discretization for quark fields on the lattice are Wilson
fermions. 
% \cite{Wilson:1974sk}
However, as a consequence of removing the unwanted doublers in the naive
lattice fermion action, it exhibits leading cutoff effects of $\Or(a)$ and
the explicit breaking of chiral symmetry.
As for the former, a systematic way of resolving this is the Symanzik 
improvement programme, 
% \cite{Symanzik:1983dc,Symanzik:1983gh} 
which amounts to add the so-called clover term to the action and further
irrelevant operators to local composite fields, canceling their $\Or(a)$
corrections, while the latter is accounted for by introducing finite
renormalization constants. 
To eliminate all $\Or(a)$ contributions from physical quantities and to
restore chiral symmetry at this order, these improvement counterterms and
renormalization factors have to be fixed non-perturbatively.

In this work we specifically look at the renormalized and improved isovector 
current, which in the chiral limit of vanishing sea quark masses and at 
non-zero valence quark mass can be parametrized as
\begin{flalign}
&& (V_\mathrm{R})_\mu^a(x)
&=
Z_\mathrm{V}(1+b_\mathrm{V}a\mq)(V_\mathrm{I})_\mu^a(x) \;,&&\\
\text{with} \nonumber\\
&& (V_\mathrm{I})_\mu^a(x)
&=
V_\mu^a(x)+ac_\mathrm{V}\tilde{\partial}_\nu T_{\mu\nu}^a(x) &&\\
&& 
&=
\overline{\psi}(x)\gamma_\mu\frac{\tau^a}{2}\psi(x)
+\mathrm{i}ac_\mathrm{V}\tilde{\partial}_\nu\overline{\psi}(x)
\sigma_{\mu\nu}\frac{\tau^a}{2}\psi(x) \;,&&
\end{flalign}
where $\tau^a$ acts in flavour space and $\tilde{\partial}_\mu$ denotes
the symmetric lattice derivative.
The quark mass dependent $\Or(a)$ improvement term proportional to 
$b_{\rm V}$, which corrects for quark mass dependent cutoff effects,
was recently calculated non-perturbatively for $\nf=3$ in 
\cite{Korcyl:2016ugy,Korcyl:2016cmx}.
The renormalization constant $\zv$ and the improvement factor $\cv$,
however, have not yet been investigated non-perturbatively so far in the
case of three-flavour QCD and are subject to this work.
Potential applications of the vector current and its matrix elements include 
computations of semi-leptonic decay form factors and of the timelike pion 
form factor, as well as contributions to the anomalous magnetic moment of 
the muon and thermal correlators related to the di-lepton production rate in 
the quark-gluon plasma.

Another object of interest is the ratio $\zs/\zp$ of the scalar and the
pseudoscalar renormalization constants, which plays a r\^{o}le in relating
renormalized PCAC and subtracted quark masses to each other.
The two constants themselves exhibit a scale dependence that cancels in the 
ratio, though.
Accordingly, in the zero sea quark mass limit and at non-vanishing valence
quark mass, the corresponding renormalized currents are defined as
\begin{align}
(S_\mathrm{R})^a(x)=
\zs(1+b_\mathrm{S}a\mq)\overline{\psi}(x)\frac{\tau^a}{2}\psi(x) \;,\quad 
(P_\mathrm{R})^a(x)=
\zp(1+b_\mathrm{P}a\mq)\overline{\psi}(x)\gamma_5\frac{\tau^a}{2}\psi(x)
\end{align}
and already comply with $\Or(a)$ improvement without any correction terms.
%
%----------------------------------------------------------------------------
\section{Renormalization and improvement conditions}
\label{sec:ren+impr}
%----------------------------------------------------------------------------
%
All improvement and renormalization conditions explained below involve the
$\Or(a)$ improved PCAC quark mass defined as
\begin{align}
m_{\mathrm{PCAC}}=
\frac{\tilde{\partial}_0\fa(x)+a\ca\partial_0^\ast\partial_0\fp(x)}
{2\fp(x)} \;,
\label{eq:conditionmpcac}
\end{align}
with standard notation for (symmetric, backward and forward) lattice
derivatives and where $\ca$ for $\Or(a)$ improved $\nf=3$ lattice QCD with 
Wilson fermions \cite{Bulava:2013cta} and tree-level improved gauge action,
% \cite{Luscher:1984xn}
as employed here, is non-perturbatively known from \cite{Bulava:2015bxa}.

Let us mention that there exists a very promising alternative approach to
determine renormalization factors through imposing appropriate conditions 
based on the PCAC relation in the Schr\"odinger functional with chirally 
rotated boundary conditions, see~\cite{Brida:2016rmy}, where it was also
tested in perturbation theory.
Apart from its advantage of entailing automatic $\Or(a)$ improvement,
it turned out that, e.g., in case of the renormalization factor of the
axial current for $\nf=2$, more precise results than with standard
Schr\"odinger functional boundary conditions can be 
obtained \cite{Brida:2014zwa}.
\subsection{Renormalization of the vector current}
The renormalization condition for the vector current is derived from the 
vector Ward identity \cite{Luscher:1996jn}, 
\begin{align}
\int_{\partial R}\mathrm{d}\sigma_\mu(x)\,
\langle V_\mu^a(x)\mathcal{O}_{\rm int}^b(y)\mathcal{O}_{\rm ext}^c(z)\rangle= 
-\,\langle
[\delta_\mathrm{V}^a\mathcal{O}_{\rm int}^b(y)]\mathcal{O}_{\rm ext}^c(z)
\rangle
\;.
\label{eq:vectorwardidentity}
\end{align}	
By choosing the spacetime region $R$ to consist of all times smaller than 
$x_0$, the only contribution stems from the timeslice $x_0$ which results 
in
\begin{align}
\int\mathrm{d}^3\mathbf{x}\,
\langle 
V_0^a(x_0,\mathbf{x})\mathcal{O}_{\rm int}^b(y)\mathcal{O}_{\rm ext}^c(z)
\rangle= 
-\,\langle
[\delta_\mathrm{V}^a\mathcal{O}_{\rm int}^b(y)]\mathcal{O}_{\rm ext}^c(z)
\rangle
\;.
\end{align}
We identify the operators $\mathcal{O}_{\rm int}$ and $\mathcal{O}_{\rm ext}$
with the boundary fields at $x_0=0$ and $x_0=T$, 
\begin{align}
\mathcal{O}^a=
a^6\sum_{\mathbf{u},\mathbf{v}}\overline{\zeta}(\mathbf{u})
\gamma_5\frac{\tau^a}{2}\zeta(\mathbf{v})
\quad \text{and} \quad
\mathcal{O}^{\prime a}=
a^6\sum_{\mathbf{u^\prime},\mathbf{v^\prime}}
\overline{\zeta}^\prime(\mathbf{u^\prime})
\gamma_5\frac{\tau^a}{2}\zeta^\prime(\mathbf{v^\prime}) \;, 
\label{eq:sourceterm}
\end{align}
where $\zeta$ and $\overline{\zeta}$ are the Schrödinger functional boundary 
fields at $x_0=0$ and their primed versions the fields at $x_0=T$,
respectively.
After replacing both sides by their renormalized lattice counterparts, 
we arrive at
\begin{align}
\zv(1+\bV a\mq)f_\mathrm{V}(x_0)=
f_1+\Or(a^2) \;,
\label{eq:conditionZV}
\end{align}
with
\begin{align}
f_\mathrm{V}(x_0)=
\frac{a^3}{2(N_f^2-1)L^6}\sum_{\mathbf{x}}\text{i}\epsilon^{abc}
\langle\mathcal{O}^{\prime c}V_0^a(x_0,\mathbf{x})\,\mathcal{O}^b\rangle
\quad \text{and} \quad
f_1=
-\frac{1}{(N_f^2-1)L^6}\,\langle\mathcal{O}^{\prime c}\mathcal{O}^{c}\rangle 
\;.
\end{align}
The Ward identity is valid for all $x_0$, although boundary effects are
expected far from the temporal center of the lattice.
In order to get a better handle on statistical fluctuations, we have
evaluated the renormalization condition at the central four timeslices and 
taken the average.
\subsection{Improvement of the vector current}
The improvement condition for the vector current was first presented in 
\cite{Guagnelli:1997db} and is based on the axial Ward identity. 
By insertion of an axial current as an operator inside the spacetime region 
$R$ we get 
\begin{align}
\int_{\partial R}\mathrm{d}\sigma_\mu(x)\,
\langle A_\mu^a(x)A_\nu^b(y)\,\mathcal{O}_{\rm ext}^c(z)\rangle
-2m\int_R\mathrm{d}^4x\,
\langle P^a(x)A_\nu^b(y)\,\mathcal{O}_{\rm ext}^c(z)\rangle=
\text{i}f^{abd}\langle V_\nu^d(y)\,\mathcal{O}_{\rm ext}^c(z)\rangle \;.
\label{eq:axialwardidentity}
\end{align}
By specifying $R$ as the region between the timeslices $x_0=t_1$ and $x_0=t_2$
with $t_1<y_0<t_2$, two surface terms arise:
\begin{align}
&
\int\mathrm{d}^3x\,
\langle[A_0^a(t_2)-A_0^a(t_1)]A_\nu^b(y) \,\mathcal{O}_{\rm ext}^c(z)\rangle
-2m\int\mathrm{d}^3\mathbf{x}\int_{t_1}^{t_2}dx_0\,
\langle P^a(x)A_\nu^b(y)\,\mathcal{O}_{\rm ext}^c(z)\rangle
\nonumber\\
&=
\,\text{i}f^{abd}\langle V_\nu^d(y)\,\mathcal{O}_{\rm ext}^c(z)\rangle \;. 
\end{align}
Since the Ward identity is valid for all $\nu$, we take $\nu=k$.
The source operator $\mathcal{O}_{\rm ext}^c$ is chosen as
\begin{align}
\mathcal{O}_k^c=
a^6\sum_{\mathbf{u},\mathbf{v}}
\overline{\zeta}(\mathbf{u})\gamma_k\frac{\tau^c}{2}\zeta(\mathbf{v}) \;,
\end{align}
where $\zeta$ and $\overline{\zeta}$ are quark fields at the boundary
$x_0=0$. 
After implementing this improvement condition in terms of Schrödinger 
functional correlation functions, one finds
\begin{align}
\za^2[k_{A_0A_k}^\mathrm{I}(t_2,y_0)-k_{A_0A_k}^\mathrm{I}(t_1,y_0)]
-2m\,\za^2\,\tilde{k}_{PA_k}(t_1,t_2,y_0)=
\zv[\kv(y_0)+a\cv\tilde{\partial}_0k_\mathrm{T}(y_0)]+\Or(a^2) \;,
\label{eq:conditioncV}
\end{align}
omitting the sea and valence quark mass $b$--coefficients for brevity;
it is to be understood as our final expression, which can be solved for
$\cv$ (once $\zv$ and $\za$ are known).
For explicit definitions of the correlators we refer, e.g., to
\cite{Luscher:1996jn,Guagnelli:1997db,DellaMorte:2005xgj,Bulava:2016ktf},
with contributions that are diagrammatically represented via quark diagrams
corresponding to possible Wick contractions in \fig{fig:ZSZPcontractions};
more details will be given elsewhere \cite{cv:toappear}.
For the present analysis, (\ref{eq:conditioncV}) was evaluated at 
%
% \begin{align}
% t_1=T/4 \;,\quad t_2=3T/4 \;, 
% \end{align}
%
$t_1=T/4$ and $t_2=3T/4$, as originally suggested in \cite{Guagnelli:1997db}.
Considered as a function of the timeslice variable $y_0$, a plateau at the
temporal center of the lattice is identified for the (local) $\cv(y_0)$. 
In order to tame statistical fluctuations, the quoted preliminary values for
$\cv$ are extracted as averages of the central two timeslices. 
\subsection{Ratio of renormalization constants $\zs/\zp$}
\begin{figure}[thb]
\sidecaption
\centering
\includegraphics[width=0.375\textwidth]{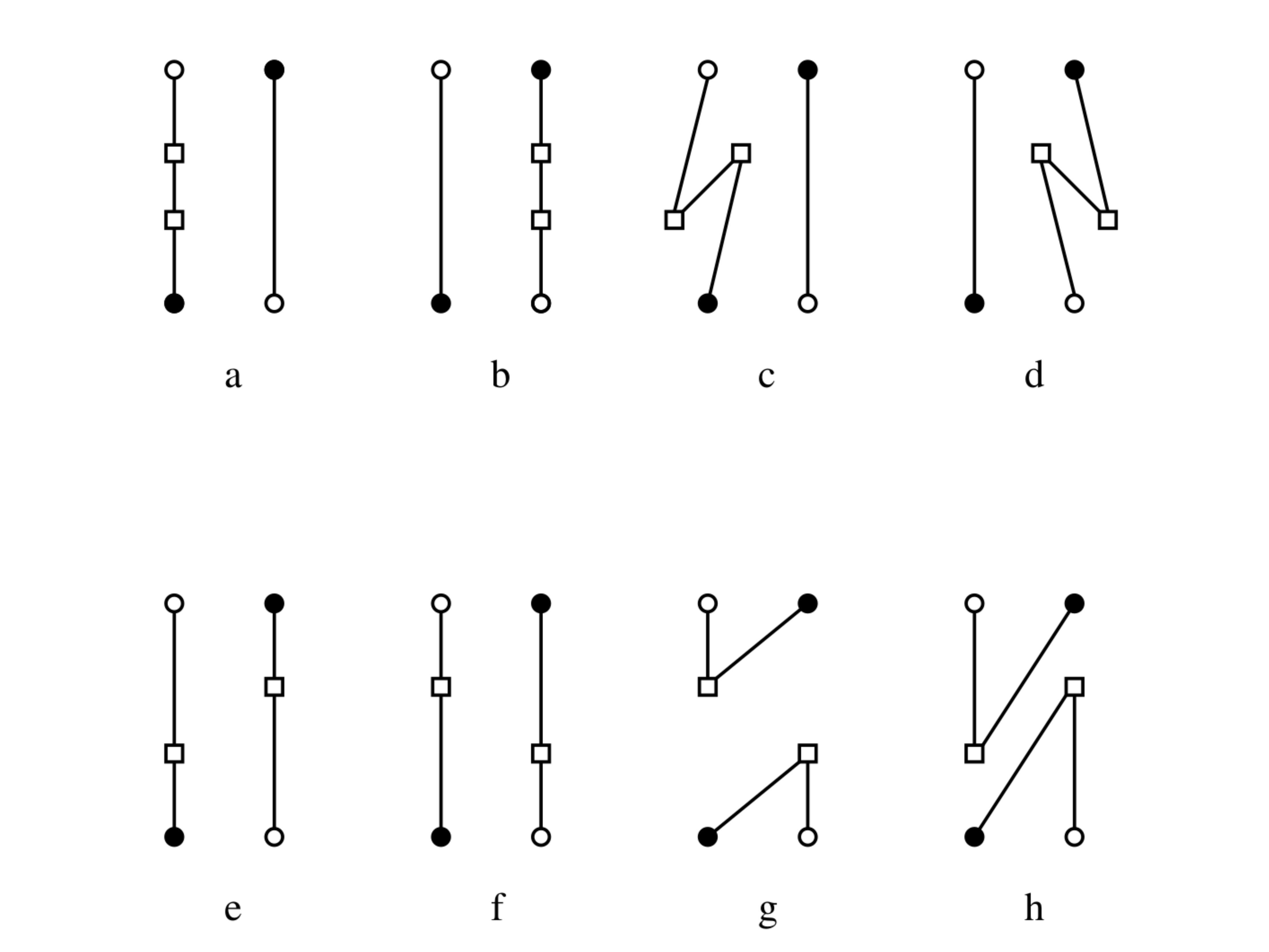}
\caption{%
Graphical representation (figure borrowed from \cite{Luscher:1996jn})
of possible Wick contractions for correlation functions of generic 
form $f_{\Gamma\,\tilde{\Gamma}}(x_0,y_0)$ with quark bilinear insertions 
$\Gamma$ and $\tilde{\Gamma}$, appearing on the l.h.s. of 
(\ref{eq:conditionZSZP}).
Filled (open) circles stand for the creation (annihilation) of a quark
at the boundaries of the lattice, while squares indicate insertions of
local composite fields.
In a first trial analysis, we had evaluated (\ref{eq:conditionZSZP}) for 
insertion points $y_0=T/2$, $t_1=T/3$ and $t_2=2T/3$.
Still, we do not quote any results for $\zp/\zs$ in this status report, 
because a careful study to extract it from the renormalization condition 
proposed here has only started after the conference.
}\label{fig:ZSZPcontractions}
\vspace{-0.625cm}
\end{figure}

To derive a renormalization condition for the ratio of quark mass
renormalization factors $\zs$ and $\zp$, we exploit a renormalization
condition that once more is derived from the massive axial Ward identity 
--- closely following the ALPHA Collaboration's method to compute $\za$ for 
$\nf=2,3$ \cite{DellaMorte:2005xgj,Bulava:2016ktf} ---, but now relying on
a pseudoscalar insertion as internal operator with a behaviour under 
variation:
\begin{align}
\delta_\mathrm{A}^aP^b(x)=
d^{abc}S^c(x)+\frac{\delta^{ab}}{\nf}\,\overline{\psi}(x)\psi(x) \;.
\label{eq:PSvariation}
\end{align}
For $d^{abc}$ not to vanish (and thus to ensure sensitivity to the scalar
density on the r.h.s. of this equation), one has to work with a 
${\rm SU}(\nf)$ algebra in the valence sector, where $\nf\geq3$.
Here, we choose ${\rm SU}(3)$, assume $a\neq b$ and adopt a product of two
pseudoscalar boundary sources, compare (\ref{eq:sourceterm}),
\begin{align}
\mathcal{O}_{\rm ext}^{ba}= 
\frac{1}{(\nf^2-1)L^6}\,\mathcal{O}^{\prime b}\mathcal{O}^a
\end{align}
as external operator in the integrated axial Ward identity
\begin{align}
\int_{\partial R}\mathrm{d}\sigma_\mu(x)\,
\langle A_\mu^a(x)\,P^b(y)\,\mathcal{O}_{\rm ext}(z)\rangle 
-2m\int_R\mathrm{d}^4x\, 
\langle P^a(x)\,P^b(y)\,\mathcal{O}_{\rm ext}(z)\rangle=
-\,d^{abc}\langle S^c(y)\,\mathcal{O}_{\rm ext}(z)\rangle \;,
\end{align}
which is similar to (\ref{eq:axialwardidentity}) but involves a pseudoscalar
density insertion with a variation according to (\ref{eq:PSvariation}).
Upon identifying each piece with a Schr\"odinger functional correlator,
some steps of algebra~\cite{zszp:toappear} yield a formula that can be
solved for $\zp/\zs$ (once $\za$ is known) and in which the intrinsic scale
dependence of the individual renormalization factors drops out, viz.
\begin{align}
\za\zp\,[\big(f_{AP}^{ba}\big)^{\rm I}(t_2,y_0)
-\big(f_{AP}^{ba}\big)^{\rm I}(t_1,y_0)-2m\,\tilde{f}_{PP}^{ba}(t_2,t_1,y_0)]=
-\,\zs f_{\rm S}^{ba}(y_0)+\Or(a^2) \;.
\label{eq:conditionZSZP}
\end{align}
Again, any $b$--coefficients are suppressed here.
The correlators are defined analogously to those in
\cite{Luscher:1996jn,Guagnelli:1997db,DellaMorte:2005xgj,Bulava:2016ktf},
and their explicit forms will be given fully elsewhere 
\cite{zszp:toappear}.
Quark diagrams with possible Wick contractions for the 
$f_{\Gamma\,\tilde{\Gamma}}(x_0,y_0)$ contributing to the l.h.s. of this
equation are illustrated in \fig{fig:ZSZPcontractions}.
%
%%%%%%%%%%%%%%%%%%%%
%%%%%%%%%%%%%%%%%%%%
%
%----------------------------------------------------------------------------
\section{Simulation details}
\label{sec:sims}
%----------------------------------------------------------------------------
%
As improvement coefficients and renormalization factors are short-distance
quantities, they can be extracted by imposing suitable conditions in
a finite (i.e., in practice, small) physical volume.
This is realized by the Schr\"odinger functional framework, governed by
periodic boundary conditions in space and Dirichlet ones in time.
The gauge field configuration ensembles used in this work are almost
identical to the ones that were generated in the context of the improvement
and renormalization of the axial vector 
current \cite{Bulava:2015bxa,Bulava:2016ktf} and cover the $\beta$--range of
the $\nf=3$ large-volume QCD configurations of the CLS effort, corresponding 
to lattice spacings of about $(0.05\lesssim a\lesssim 0.09)\,\Fm$ 
\cite{Bruno:2014jqa,Bruno:2016plf}.
Their specifications are collected in \tab{tab:simparams}.
%
%%%%%%%%%%%%%%%%%%%%
\begin{table}[thb]
\sidecaptionMy
\small
\centering
\begin{tabular}{cllrrc}
\toprule
$L^3\times T/a^4$ & $\beta$ & $\kappa$ & \#REP & \#MDU & ID   \\
\midrule
$12^3\times17$ & 3.3   & 0.13652 & 20 & 10240 & A1k1          \\
               &       & 0.13660 & 10 & 13672 & A1k2          \\
               &       & 0.13648 &  5 &  6876 & \textbf{A1k3} \\
\midrule
$14^3\times21$ & 3.414 & 0.13690 & 32 & 10176 & E1k1          \\
               &       & 0.13695 & 48 & 13976 & E1k2          \\
\midrule
$16^3\times23$ & 3.512 & 0.13700 &  2 & 20480 & B1k1          \\
               &       & 0.13703 &  1 & 8192  & B1k2          \\
               &       & 0.13710 &  3 & 22528 & B1k3          \\
\midrule
$16^3\times23$ & 3.47  & 0.13700 &  3 & 29560 & B2k1          \\
\midrule
$20^3\times29$ & 3.676 & 0.13700 &  4 & 15232 & C1k2          \\
               &       & 0.13719 &  4 & 15472 & C1k3          \\
\midrule
$24^3\times35$ & 3.810 & 0.13712 &  6 & 10272 & D1k1          \\
               &       & 0.13701 &  3 &  5672 & \textbf{D1k2} \\
               &       & 0.13704 &  1 &   800 & \textbf{D1k3} \\
\bottomrule
\end{tabular}
\caption{%
Summary of simulation parameters of the gauge configuration ensembles used
in this study, as well as the number of (statistically independent) replica
per ensemble `ID' and their total number of molecular dynamics units.
Bold ID's indicate three new ensembles compared 
to \cite{Bulava:2015bxa,Bulava:2016ktf}.
}\label{tab:simparams}
\end{table}
%%%%%%%%%%%%%%%%%%%%
%
To supplement the data base of configurations already available
from \cite{Bulava:2015bxa,Bulava:2016ktf}, the production of a few new
ensembles --- labeled by A1k3, D1k2 and D1k3 in the table --- was started.
These ensembles exhibit a more chiral (i.e., closer to zero) mass of the
three mass-degenerate sea quarks and thereby allow for getting a better
handle on the mass dependencies of the quantities of interest that will
prove to be essential in the case of $\cv$. 

Compared to the previous $\nf=0$ study \cite{Guagnelli:1997db},
we have implemented various refinements:
First of all, as detailed in \cite{Bulava:2015bxa}, all gauge field 
ensembles entering the analysis lie on a line of constant physics
characterized by a fixed spatial physical volume of
$L\approx 1.2\fm={\rm constant}$, $T\approx 3L/2$ and almost vanishing
mass of the (degenerate) sea quarks.
The valence quark mass in the computation of correlation functions equals
the sea quark value. 
This entails that the renormalization and improvement factors become
smooth functions of the bare coupling, i.e., $g_0^2=6/\beta$.
Only the ensemble B2k1 deliberately misses the condition of fixed physical
volume and is used to quantitatively investigate the effect of such 
a deviation on the results. 
Furthermore, again following \cite{Bulava:2015bxa}, the Schrödinger 
functional correlation functions incorporate optimized boundary 
wave-functions, in order to suppress excited state effects and thus to
maximize the overlap with the ground state in their spectral decomposition.
Finally, for the case of $\cv$, we also have identified the importance of
the additional mass term in the axial Ward identity, (\ref{eq:conditioncV}), 
the impact of which will be discussed in the results section.
In this context, we have tested different sets of insertion times for the
individual operators and found a specific choice that seems to reduce the
effects caused by the non-zero mass comprehensively.

The statistical error analysis of the Markov chain Monte Carlo data 
utilizes the $\Gamma$--method based on evaluating autocorrelation functions
\cite{Wolff:2003sm} and was cross-checked against binned Jackknife
estimates.
%
%----------------------------------------------------------------------------
\section{Results}
\label{sec:res}
%----------------------------------------------------------------------------
%
The analysis underlying the results presented here was done including all
topological sectors.
By virtue of the theoretical argument that our results --- being based on
Ward identities, as operator identities holding in any topological 
sector --- should be insensitive to the topological charge $\qtop$, 
we believe that the influence of restricting the computations to one sector
of fixed $\qtop$ (say, $\qtop=0$) is negligible, modulo the accompanying
reduction of statistics.
This expectation still needs to be confirmed in the final analysis, though. 

\begin{figure}[!htb]
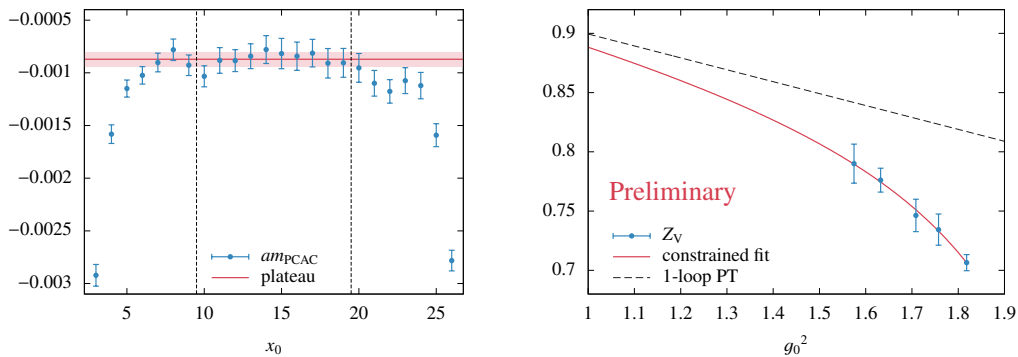

\centering
\begin{minipage}{0.497\textwidth}
\centering
\graphicspath{{figures/mplateau/}}
\resizebox{\textwidth}{!}{\input{figures/mplateau/mplateautex}}
\end{minipage}
\begin{minipage}{0.497\textwidth}
\centering
\graphicspath{{figures/ZV/}}
\resizebox{\textwidth}{!}{\input{figures/ZV/ZVtex}}
\end{minipage}
\vspace{-0.25cm}
\caption{%
{\sl Left:}
Exemplary results for $am_{\mathrm{PCAC}}$ on different timeslices,
evaluated on the C1k3 ensemble.
The dashed vertical lines span the plateau region of the central $L/2$ 
timeslices, over which the plateau average is taken.
{\sl Right:}
Results for $\zv$ together with an interpolating fit,
constrained by 1--loop perturbation theory \cite{Aoki:1998ar}.
}\label{fig:mPCAC+ZVfirstresult}
\vspace{-0.75cm}
\end{figure}

The left panel of \fig{fig:mPCAC+ZVfirstresult} shows an representative
evaluation of the local PCAC quark mass, $am_{\mathrm{PCAC}}(x_0)$, on the
gauge configurations of ensemble C1k3.
The actual values for all PCAC masses entering our analysis are always
chosen as plateau averages over the central $L/2$ timeslices of the 
temporal extent of the lattice.
Thereby it is guaranteed that also these averaging intervals are scaled in
physical units in the same way as all other length scales, in order to obey
the constant physics condition in all steps of the computation.
%
%%%%%%%%%%%%%%%%%%%%
%%%%%%%%%%%%%%%%%%%%
%
\subsection{$\zv$}
Results for the vector renormalization constant $\zv$ are presented in the 
right panel of \fig{fig:mPCAC+ZVfirstresult}, in comparison to 1--loop 
perturbation theory taken from \cite{Aoki:1998ar}.
The individual data points, obtained by a chiral extrapolation to zero
(valence $=$ sea) quark mass using $\bV$ in (\ref{eq:conditionZV}) from
\cite{Korcyl:2016ugy,Korcyl:2016cmx} (but neglecting the corresponding 
$b$--coefficient in the sea quark sector), show a smooth behavior that is 
well described by a polynomial fit constrained by perturbation theory.
The preliminary interpolation formula reads:
\begin{align}
\zv(g_0^2)=
1-0.10057g_0^2\times\frac{1-0.388(13)g_0^2}{1-0.449(8)g_0^2} \;.
\label{eq:ZVinterpolationformula}
\end{align}
\subsection{$\cv$}
%
%%%%%%%%%%%%%%%%%%%%
\begin{figure}[!htb]
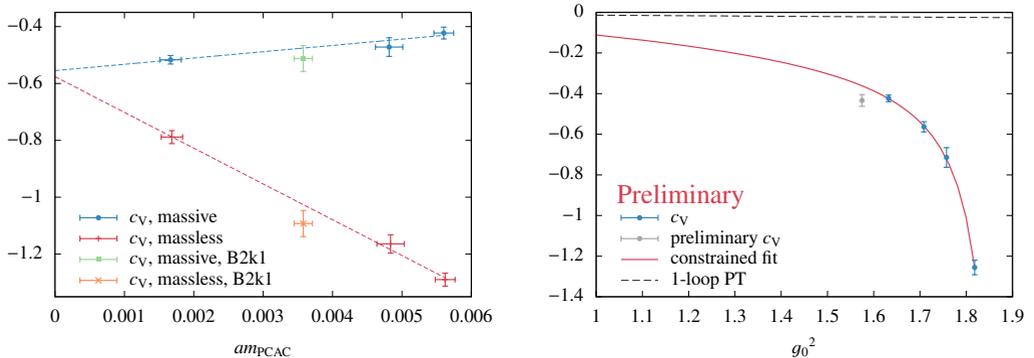

\centering
\begin{minipage}{0.497\textwidth}
\centering
\graphicspath{{figures/cVchiral/}}
\resizebox{\textwidth}{!}{\input{figures/cVchiral/cVchiraltex}}
\end{minipage}
\begin{minipage}{0.497\textwidth}
\centering
\graphicspath{{figures/cV/}}
\resizebox{\textwidth}{!}{\input{figures/cV/cVtex}}
\end{minipage}
\vspace{-0.25cm}
\caption{%
{\sl Left:}
Chiral extrapolation of $\cv$ at $L/a=16$ and $g_0^2=1.7084$ ($\beta=3.512$) 
for the case with and without including the additional mass term in the Ward 
identity.
The data points from the ensemble B2k1 indicate that a sizeable violation of 
the constant physics condition does not influence our results appreciably.
{\sl Right:}
Results for $\cv$ together with an interpolating fit,
constrained by 1--loop perturbation theory \cite{Taniguchi:1998pf}.
The gray data point refers to a tentative analysis on the gauge 
configuration ensembles D1k2 and D1k3, whose generation was launched
during the conference.
Hence, it is only indicative and not yet included in the fit.
}\label{fig:cVchiral+cVfirstresult}
\vspace{-0.5cm}
\end{figure}
%%%%%%%%%%%%%%%%%%%%
%
As outlined above, the condition of \cite{Guagnelli:1997db} to fix the
vector current improvement coefficient $\cv$ is extended by accounting for 
an additional term that naturally arises when the Ward identity is employed
at finite quark mass, cf. (\ref{eq:conditioncV}).
The impact of this term is demonstrated in the left panel 
of \fig{fig:cVchiral+cVfirstresult}, where the chiral extrapolations 
(using values for the associated valence quark mass $b$--coefficients from 
\cite{Korcyl:2016ugy,Korcyl:2016cmx}) for $L/a=16$ and $g_0^2=1.7084$ 
($\beta=3.512$) with and without the mass term are compared with each other. 
Although both extrapolations nicely meet in almost the same chiral limit
at $am_{\mathrm{PCAC}}=0$, the data without inclusion of the mass term show
a much steeper behaviour.
This finally results in a larger error at $am_{\mathrm{PCAC}}=0$ and thus
underlines the importance of refining the improvement condition for $\cv$
through accounting for the mass term in the analysis even at small but 
finite quark masses.

Additionally, to quantitatively check for the influence of a violation of
the constant physics condition on our analysis, results from the gauge
configuration ensemble B2k1 are displayed in the same figure.
This ensemble (see \tab{tab:simparams}) has a $\beta$--value shifted by an
amount, which corresponds to a $\sim 6\%$ shift in the spatial extent of
the physical volume, and thus induces a significant deviation from our
condition $L\approx 1.2\fm={\rm constant}$.   
As can be seen in the left panel of \fig{fig:cVchiral+cVfirstresult},
these data points align fairly well with the other points along the 
(linear) fit function.
Hence, we conclude that any deviations from the constant physics condition
of this order of magnitude or below are of only minor influence and can
safely be neglected on the level of the final precision for 
$\cv$.\footnote{%
The same holds true for our results on $\zv$.
}
Note that this is also in line with the findings already reported in
\cite{Bulava:2015bxa,Bulava:2016ktf}.

The preliminary estimates for $\cv$ are presented in the right panel of
\fig{fig:cVchiral+cVfirstresult}, together with the prediction from 1--loop
perturbation theory that we have extracted for our lattice action from the
perturbative results in \cite{Taniguchi:1998pf}.
The gray data point stems from the ensembles D1k2 and D1k3 
(see \tab{tab:simparams}).
Since the generation of these gauge field configurations was only started
during the conference and is still ongoing, we exclude it from the 
subsequent analysis steps for the purpose of the present status report.
Nevertheless it is reassuring that this --- so far only indicative ---
result appears to blend in well with the $g_0^2$--dependence of the other
points.

At this point, we therefore describe our results for $\cv$ by a
preliminary interpolating Pad\'e fit, constrained by 1--loop perturbation
theory in the asymptotic $g_0^2\to 0$ regime, as
\begin{align}
\cv(g_0^2)=
-0.01030(4)g_0^2C_{\rm F}\times
\frac{1+5.80(47)g_0^2-2.99(30)g_0^4}{1-0.532(1)g_0^2} \;,\quad 
C_{\rm F}=\frac{4}{3} \;,
\label{eq:cVinterpolationformula}
\end{align}
where, as stressed above, the gray point in the right plot of
\fig{fig:cVchiral+cVfirstresult} is not included in the fit.
Moreover, any uncertainties originating from $\zv$ or $\za$ (entering the
final formula for $\cv$ according to (\ref{eq:conditioncV})) have not
yet been propagated into the errors on $\cv$ quoted in the figure such that
we still expect them to slightly increase after a final analysis.
Note that, in qualitative agreement with observations already made in the
exploratory quenched study \cite{Guagnelli:1997db}, the non-perturbative 
$\cv$ substantially deviates from perturbation theory in the range of
bare couplings (resp. $\beta$--values) typically encountered in
large-volume applications with the lattice action employed here.
%
%----------------------------------------------------------------------------
\section{Outlook}
\label{sec:outl}
%----------------------------------------------------------------------------
%
For the completion of our work to determine the renormalization and
improvement factors discussed in this report it essentially remains to
1.) evaluate the relevant correlators for the full statistics on all
ensembles of \tab{tab:simparams},
2.) check for independence of the results on topology by repeating the
computations in the sectors of fixed $\qtop$, 
3.) quantify the size of possible $\Or(a)$ ambiguities in improvement
(resp. renormalization) conditions for the vector current and, in particular,
4.) to also perform the data analysis to extract the ratio $\zs/\zp$.
For a related study to calculate improvement $b$--coefficients in the
valence sector, multiplying mass dependent $\Or(a)$ Symanzik counterterms
to local operators, as well as the ratio $\zm\zp/\za$ of quark mass 
renormalization constants, see \cite{deDivitiis:2017vvw}.
%
%%%%%%%%%%%%%%%%%%%%%%%%%%%%%%%%%%
% acknowledgments and bibliography
%%%%%%%%%%%%%%%%%%%%%%%%%%%%%%%%%%
%
\subsection*{Acknowledgements}
%
% \begin{acknowledgement}
We thank P.~Fritzsch, S.~Sint, R.~Sommer, S.~Kuberski and G.~Bali 
for helpful discussions.
This work was supported by 
% the 
grant {HE 4517/3-1} (J.~H. and C.~W.)
of the {\it Deutsche Forschungsgemeinschaft}.
We gratefully acknowledge the computer resources provided by the
{\it Zentrum für Informationsverarbeitung} of the University of
M\"{u}nster (PALMA HPC cluster) and thank its staff for 
% their 
support.
% \end{acknowledgement}
%
% \clearpage
%
% BibTeX or Biber users please use 
% (since the style is already called in the class,
% ensure that the "woc.bst" style is in your local directory):
% \bibliography{name or your bibliography database}
%
\bibliography{lattice2017}
%
%%%%%%%%%%%%%%%%%%%%%%%%%%%%%%%%%%%%%%%%%%%%%%%%%%%%%%%%%%%%%%%%%%%%%%%%%%%%%
\end{document}